# The anachronism of whole-GPU accounting

Igor Sfiligoi

University of California San Diego, La Jolla, CA, USA, isfiligoi@sdsc.edu

David Schultz

University of Wisconsin-Madison, Madison, WI, USA, david.schultz@icecube.wisc.edu

Frank Würthwein

University of California San Diego, La Jolla, CA, USA, fkw@ucsd.edu

Benedikt Riedel

University of Wisconsin-Madison, Madison, WI, USA, benedikt.riedel@icecube.wisc.edu

Dmitry Y. Mishin

University of California San Diego, La Jolla, CA, USA, dmishin@ucsd.edu

NVIDIA has been making steady progress in increasing the compute performance of its GPUs, resulting in order of magnitude compute throughput improvements over the years. With several models of GPUs coexisting in many deployments, the traditional accounting method of treating all GPUs as being equal is not reflecting compute output anymore. Moreover, for applications that require significant CPU-based compute to complement the GPU-based compute, it is becoming harder and harder to make full use of the newer GPUs, requiring sharing of those GPUs between multiple applications in order to maximize the achievable science output. This further reduces the value of whole-GPU accounting, especially when the sharing is done at the infrastructure level. We thus argue that GPU accounting for throughput-oriented infrastructures should be expressed in GPU core hours, much like it is normally done for the CPUs. While GPU core compute throughput does change between GPU generations, the variability is similar to what we expect to see among CPU cores. To validate our position, we present an extensive set of run time measurements of two IceCube photon propagation workflows on 14 GPU models, using both on-prem and Cloud resources. The measurements also outline the influence of GPU sharing at both HTCondor and Kubernetes infrastructure level.

## 1 INTRODUCTION

General purpose Graphics Processor Unit (GPU) accelerated computing has moved from being a niche use case to being considered mainstream in many domains, including Machine Learning (ML) and High Performance Computing (HPC) simulation environments. The IceCube Neutrino Observatory [1] detector simulation workflows have been one of the early adopters of GPU accelerated compute and remain the leading consumer of GPU resources inside the Open Science Grid (OSG) [2] environment. The amount of simulation output, aka throughput, IceCube is able to get in a day from a single GPU has been steadily increasing with each new generation of NVIDIA GPUs, but older GPU models remain an important contributor to the total science output. There is thus a growing desire to proportionally account the contributed compute resources based on the GPU model, and not just use "*GPU hours*" as the main metric.

IceCube's internal accounting system has been dealing with GPU heterogeneity by running dedicated benchmarks every time a new GPU model was encountered, and thus scaling the contribution based on that [3]. While very reliable for their use case, a more generic accounting solution is needed for general purpose GPU-providing infrastructure providers, like the OSG and the Pacific Research Platform (PRP) [4]. Cloud providers may benefit from it, too, although cost-based accounting there provides a viable alternative already.

The accounting situation is further complicated by the need for GPU sharing, especially at the infrastructure level. As GPU models become faster and faster, it is becoming increasingly hard for an application to keep the GPU fully utilized, since most applications drive the GPU compute from a set of CPU cores. Mapping multiple compute jobs on the same GPU can thus result in a significant boost in total throughput. Each job will however take longer to complete, so it is desirable to properly account for such sharing.

In this paper we explore the option of using "*GPU core hours*" as a fair metric for GPU accounting for throughput-oriented infrastructures, much like "*CPU core hours*" has been the main metric used by CPU-focused accounting systems. To validate our assumptions, we measured the runtimes for two IceCube workflows, representing the two extremes of GPU-to-CPU compute ratio, on 14 different GPU models spanning both on-prem and Cloud setups. In the process we also show the importance of GPU sharing to maximize the achievable throughput.

Section 2 provides an introduction to both the two IceCube workflows and the infrastructure setups used for the measurement, alongside the summary of measured runtimes and how they would have been accounted using the standard "*GPU hours*" metric. In Section 3 we analyze those results and showcase the benefits of GPU sharing. Finally, in Section 4 we correlate the observed throughput with the hardware characteristics of the used GPUs, i.e. the GPU core counts, establishing that the "*GPU core hours*" is indeed a reasonable accounting metric for GPU resources.

## 2 TEST SETUP AND RAW RESULTS

The IceCube detector is located at the geographical south pole and is embedded into the naturally occurring ice there. In order to properly characterize the optical properties of that ice, IceCube has to perform extensive simulation. The problem is too complex for a parametrized approach, so brute-force photon propagation, aka ray-tracing is used. The massive, pleasantly parallel nature of the problem fits perfectly the GPU compute model [5] so all such compute has long been performed exclusively on GPU resources. Nevertheless, the initial properties of the tracked photon are provided from CPU-driven code, which is only minimally parallelizable and cannot effectively use more than two CPU cores, and that can become a bottleneck as GPU compute returns results faster and faster. Not all problems need the same amount of detail, so IceCube users can choose to speed up their compute by, roughly speaking, increasing the size of the target for the photons by some factor, with `oversize=1` being the most precise and `oversize=4` being the fastest.

The collected runtimes belong to two production IceCube workflows, one using `oversize=1` and one using `oversize=4`, filtering out any jobs that did not run on the resources under our direct control. For the purpose of this paper, we limited ourselves to the GPU resources managed by the PRP Kubernetes on-prem cluster and a dedicated Google Kubernetes Engine (GKE) Cloud-based cluster. Since IceCube uses HTCondor for job scheduling, a HTCondor pilot [6] was first scheduled on the Kubernetes-managed resources, and HTCondor would then launch the appropriate IceCube job. Each IceCube workflow is composed of thousands of independent simulation jobs and each job operates on slightly different inputs, so runtimes will vary stochastically even on the same identical hardware. We thus group the data by workflow and GPU type, and provide the mean and standard deviation values as the representative measurements.

A fraction of the provisioned GPUs were shared between multiple applications, either at infrastructure or pilot level, since the jobs used less than 5GB of GPU memory each. At the infrastructure level, we tested hardware GPU sharing for the A100 GPUs on both PRP and GKE, using the NVIDIA Multi-Instance GPU (MIG) capability [7], as well as temporally multiplexed GPU sharing on GKE. Inside the HTCondor-based pilots, the GPU sharing was achieved by configuring multiple execution slots that all pointed to the same GPU, which is conceptually comparable to the GKE approach. Note that the infrastructure and pilot level sharing can be combined, and we did run a fraction of the jobs in such a setup.

Table 1: Measured values for IceCube jobs grouped by resource provider, GPU model and sharing setup. *Job runtime* is the mean value of the measured runtimes, in kseconds, and *Jobs per unit* is the mean of the seconds_in_unit/runtime transformation.

| Infrastructure | GPU model | MIG Partitions | Kubernetes Sharing | HTCondor sharing | Job runtime oversize=1 | Jobs per day oversize=1 | Job runtime oversize=4 | Jobs per hour oversize=4 |
|---|---|---|---|---|---|---|---|---|
| GKE | Tesla K80 | None | None | None | | | 1.82±0.15 | 2.0±0.2 |
| GKE | Tesla T4 | None | None | None | 11.5±0.2 | 7.5±0.1 | 0.87±0.08 | 4.2±0.4 |
| GKE | Tesla T4 | None | 2x | None | 22.5±0.9 | 7.1±0.3 | 1.18±0.19 | 6.2±1.1 |
| GKE | Tesla T4 | None | None | 2x | | | 1.15±0.15 | 6.4±0.8 |
| GKE | V100-SXM2 | None | None | None | 7.8±0.3 | 11.0±0.3 | 0.88±0.08 | 4.1±0.4 |
| GKE | V100-SXM2 | None | 2x | None | 14.2±0.2 | 12.1±0.2 | 1.06±0.13 | 6.9±0.9 |
| GKE | V100-SXM2 | None | 3x | None | | | 1.43±0.15 | 7.7±0.9 |
| GKE | V100-SXM2 | None | None | 3x | | | 1.34±0.14 | 8.2±0.9 |
| GKE | A100-SXM4 | None | None | None | 7.3±0.1 | 11.8±0.2 | 0.83±0.10 | 4.3±0.4 |
| GKE | A100-SXM4 | 2x | None | None | 9.0±0.1 | 19.3±0.1 | 0.91±0.08 | 8.0±0.7 |
| GKE | A100-SXM4 | None | 2x | None | 8.4±0.1 | 20.7±0.2 | 0.86±0.08 | 8.4±0.7 |
| GKE | A100-SXM4 | None | None | 2x | | | 0.78±0.06 | 9.2±0.7 |
| GKE | A100-SXM4 | 2x | None | 3x | | | 1.49±0.07 | 14.6±0.7 |
| GKE | A100-SXM4 | 7x | None | None | 26.1±0.2 | 23.2±0.1 | 1.39±0.12 | 18.2±1.5 |
| GKE | A100-SXM4 | None | 7x | None | 26.8±0.1 | 22.6±0.1 | 1.32±0.12 | 19.3±1.7 |
| GKE | A100-SXM4 | None | None | 7x | | | 1.34±0.12 | 19.0±1.8 |
| PRP | Quadro M8000 | None | None | None | 43.7±0.3 | 2.0±0.0 | 1.95±0.16 | 1.9±0.2 |
| PRP | GTX 1070 | None | None | None | 16.4±0.2 | 5.2±0.1 | 1.04±0.09 | 3.5±0.3 |
| PRP | GTX 1080 | None | None | None | 12.4±0.4 | 7.0±0.2 | 0.76±0.19 | 4.8±0.5 |
| PRP | GTX 1080 Ti | None | None | None | 9.0±0.5 | 9.7±0.4 | 0.76±0.09 | 4.9±1.0 |
| PRP | RTX 2080 Ti | None | None | None | 6.4±0.4 | 13.7±0.8 | 0.74±0.09 | 4.9±0.6 |
| PRP | RTX 2080 Ti | None | None | 2x | | | 1.23±0.19 | 6.0±0.9 |
| PRP | Titan RTX | None | None | None | 5.9±0.5 | 14.7±1.1 | 0.66±0.17 | 5.7±1.1 |
| PRP | V100-SXM2 | None | None | None | 6.2±0.1 | 13.9±0.1 | 0.61±0.05 | 5.9±0.5 |
| PRP | V100-SXM2 | None | None | 3x | | | 1.23±0.18 | 9.0±1.4 |
| PRP | A100-PCIE | None | None | None | 4.5±0.1 | 19.3±0.6 | 0.58±0.05 | 6.2±0.6 |
| PRP | A100-PCIE | 2x | None | None | 9.4±0.1 | 18.4±0.1 | 0.66±0.05 | 11.1±0.9 |
| PRP | A100-PCIE | 3x | None | None | | | 0.84±0.07 | 12.9±1.0 |
| PRP | A100-PCIE | 3x | None | 2x | | | 1.35±0.14 | 16.2±1.8 |
| PRP | A40 | None | None | None | 4.2±0.2 | 20.9±1.3 | 0.55±0.06 | 6.6±0.6 |
| PRP | A40 | None | None | 4x | | | 0.85±0.09 | 17.1±1.9 |
| PRP | RTX 3090 | None | None | None | 3.9±0.4 | 22.4±1.8 | 0.65±0.05 | 5.6±0.5 |
| PRP | RTX 3090 | None | None | 4x | | | 0.85±0.09 | 17.1±2.0 |

The measured runtime results for the two workflows are available in Table 1. Before analyzing the results, we would like to stress that while some of the GPU models in this list are indeed old, they are not obsolete. For example, for throughput-oriented workloads the Tesla K80 GPU is still one of the most cost-effective GPUs that one can rent in the Google Cloud [8], at just $0.91/day vs 17.76/day for a V100-SXM2, in spot mode.

When looking at the measured runtimes, one can make two straightforward observations:

1. The number of jobs per unit of time varies by a factor 10x between GPU models, for both workloads. It is thus obvious that accounting only in "*GPU hours*" will not provide a reasonable estimate of the delivered value, i.e. completed science jobs, in shared heterogeneous deployments.
2. The minimum number of applications that must share a GPU to maximize its throughput varies by a factor 7x between GPU models, and it depends both on the workload and supporting environment; for example, the A100 GPUs in PRP are paired with high-frequency low-core-count CPUs, reported as "AMD EPYC 7252 8-Core Processor", while the A100 GPUs in GKE are paired with low-frequency high-core-count CPUs, reported as "Intel(R) Xeon(R) CPU @ 2.20GHz" model 85 stepping 7, resulting in a very different optimal sharing factor for the two. That said, not all applications may be willing, or capable, to share at the maximum ratio, either due to GPU memory sizes or maximum acceptable runtimes. The optimal GPU sharing strategy will thus likely be determined by a mix of infrastructure and user needs, and is likely to change with time. Naively summing the consumed "GPU hours" for each completed job for GPU accounting purposes will thus provide a highly inaccurate picture of the delivered value.

## 3 THE BENEFITS OF GPU SHARING

As seen in Table 1, many workloads cannot directly make effective use of recent GPU models and sharing a GPU between multiple applications becomes necessary to maximize the achievable throughput. In IceCube case, `oversize=4` workflow reaches that point already with the low-power Tesla T4, but even the `oversize=1` workload will benefit from sharing of the high-end A100 GPU, when paired with a high-core-count but low-frequency CPU. The benefits of GPU sharing of course change with GPU model and application type, as summarized in Figure 1, but we do observe a 4.5x speedup in the best case, i.e. the `oversize=4` workflow running on GKE A100 GPUs.

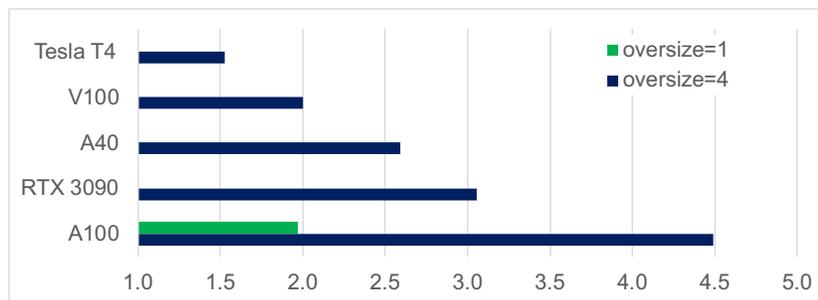

Figure 1: Ratio of observed *jobs per unit of time*, between best-option shared GPU and full GPU setups.

Choosing the right layer for GPU sharing is mostly a policy issue. The observed runtimes between the three options are roughly comparable in all the scenarios we ran in. The hardware partitioning provides the best isolation properties and would likely be the preferred option in multi-tenant deployments, but it is quite rigid [7] and only supported on the A100 GPU. Direct infrastructure level GPU sharing, either using Kubernetes as we did, or HTCondor in batch-managed deployments, is much more flexible but provides very limited isolation guarantees. Pilot level sharing using HTCondor is conceptually similar to infrastructure sharing, but comes with increased draining waste [9] due to the larger job slot count. For completeness, it should be noted that not all user communities rely on a pilot scheduling layer.

## 4 USING GPU CORE HOURS AS A GPU ACCOUNTING METRIC

Having noted that "*GPU hour*" accounting does not provide a reliable picture of delivered value, we propose an alternative metric, i.e. the simple but reliable "*GPU core hours*" metric, similarly to how "*CPU core hours*" is one of the most popular CPU accounting metrics. Domain-specific metrics would likely be the more precise, but general-purpose infrastructures typically demand application-agnostic accounting metrics. We would also like to emphasize that, to the best of our knowledge, there is no other single standardized GPU-based metric that everyone agrees on.

The number of GPU cores can easily be determined at runtime using the appropriate vendor API. Most GPU-aware resource management systems will have this capability, and we used HTCondor's `condor_gpu_discover` tool to collect the GPU core counts presented in Table 2. The same table also contains the full-utilization throughput measured, from Table 1, alongside the throughput to GPU core count ratio.

Table 2: Measured values correlated with GPU core counts.

| GPU model | Compute Units (CUs) | Cores per CU | GPU cores | Jobs per GPU oversize=1 | Jobs per kcores oversize=1 | Jobs Per GPU oversize=4 | Jobs per kcores oversize=4 |
|---|---|---|---|---|---|---|---|
| Quadro M8000 | 13 | 128 | 1664 | 2.0±0.0 | 1.2±0.0 | 1.9±0.2 | 1.14±0.12 |
| Tesla K80 | 13 | 192 | 2496 | | | 2.0±0.2 | 0.80±0.08 |
| GTX 1070 | 15 | 128 | 1920 | 5.2±0.1 | 2.7±0.1 | 3.5±0.3 | 1.82±0.16 |
| GTX 1080 | 20 | 128 | 2560 | 7.0±0.2 | 2.7±0.1 | 4.8±0.5 | 1.88±0.20 |
| GTX 1080 Ti | 28 | 128 | 3584 | 9.7±0.4 | 2.7±0.1 | 4.9±1.0 | 1.37±0.28 |
| RTX 2080 Ti | 68 | 64 | 4352 | 13.7±0.8 | 3.2±0.2 | 6.0±0.9 | 1.38±0.21 |
| Titan RTX | 72 | 64 | 4608 | 14.7±1.1 | 3.2±1.2 | 5.7±1.1 | 1.24±0.24 |
| Tesla T4 | 40 | 64 | 2560 | 7.5±0.1 | 2.9±0.0 | 6.2±1.1 | 2.42±0.43 |
| V100-SXM2 | 80 | 64 | 5120 | 13.9±0.1 | 2.7±0.0 | 9.0±1.4 | 1.76±0.27 |
| RTX 3090 | 82 | 128 | 10496 | 22.4±1.8 | 2.1±0.2 | 17.1±2.0 | 1.63±0.19 |
| A40 | 84 | 128 | 10752 | 20.9±1.3 | 1.9±0.1 | 17.1±1.9 | 1.59±0.18 |
| A100-PCIE | 108 | 64 | 6912 | 19.3±0.6 | 2.8±0.1 | 16.2±1.8 | 2.34±0.26 |
| A100-SXM4 | 108 | 64 | 6912 | 23.2±0.1 | 3.4±0.0 | 19.3±1.7 | 2.79±0.25 |
| A100 MIG 1g.5gb | 14 | 64 | 896 | 3.3±0.0 | 3.7±0.0 | 2.6±0.2 | 2.90±0.24 |
| A100 MIG 2g.10gb | 28 | 64 | 1792 | 6.5±0.0 | 3.6±0.0 | 5.4±0.6 | 3.00±0.33 |

The correlation between application throughput and GPU cores is of course not perfect. Different generations of GPUs, and even different target markets, will result in a different balance of resources inside each GPU core, and thus different application throughput. Nevertheless, the spread between highest and lowest throughput per core is only about 3x, significantly lower than the 10x spread we saw when looking at whole-GPU throughput. It is also consistent with what one typically sees when comparing throughput per CPU core.

Furthermore, when using hardware GPU partitioning, the application is aware of how many GPU cores are available to it and can properly record that information. The accounting can then be done by simply summing the attributes in the job history. In the case of GPU sharing, the situation is a little more complicated but, again, similar to what happens when a CPU is shared; most of the time an application will only get a subset of the available cores and will be accounted for accordingly. Unfortunately, at the time of writing we are not aware of any standard way to convey that information to the application, e.g. there is no GPU equivalent of OMP_NUM_THREADS, so agreeing on a standard mechanism will require some additional work. Note that similar considerations apply to allocated GPU memory accounting, too.

## 5 SUMMARY AND CONCLUSIONS

One of the main aims of any accounting system is to provide a measure of delivered value over time, which for most batch-oriented science workloads translates to the number of completed jobs per unit of time. This paper shows that the traditional whole-GPU accounting, i.e. using "*GPU hours*" as the main metric, does not currently meet that goal, due to both dramatic performance differences between GPU models and the emerging trend of GPU sharing. We back our claim by providing an extensive set of IceCube benchmark results on 14 GPU models, in both on-prem and Cloud deployments.

We propose the use of "*GPU core hours*" as a much fairer accounting metric, instead. While not perfect, we show that GPU-core counts correlate significantly better to actual application throughput than whole-GPU counts. The proposed metric also works much nicer with hardware GPU partitioning, e.g. NVIDIA A100 MIG, removing any partitioning artifacts from the accounting data.

Properly accounting for GPU sharing is, however, currently somewhat harder, due to a lack of standard sharing information propagation. Nevertheless, the problem is very similar to the one encountered by CPU-only accounting systems since most CPU chips are shared between multiple applications in throughput-oriented infrastructures. It is thus our belief that the problem is solvable, but it will require some standardization effort.

Given the increased reliance on GPU compute in throughput-oriented infrastructures, it is becoming imperative for the GPU accounting systems to match the accuracy of CPU-only accounting counterparts. We argue that this can be achieved by switching from a whole-GPU to a GPU-core-based accounting metric.

**ACKNOWLEDGMENTS**

This work has been partially funded by the US National Science Foundation (NSF) Grants OAC-1826967, OAC-2030508, CNS-1925001, OAC-1841530, CNS-1730158, OAC-2112167, CNS-2100237, CNS-2120019, OPP- 2042807 and OAC-2103963. All Google Kubernetes Engine costs have been covered by Google-issued credits.

**REFERENCES**

[1] M. G. Aartsen et al. 2015. The IceProd framework: Distributed data processing for the IceCube neutrino observatory. Journal of Parallel and Distributed Computing 75 198-211. ISSN 0743-7315. https://doi.org/10.1016/j.jpdc.2014.08.001
[2] Ruth Pordes et al. 2007. The open science grid. J. Phys.: Conf. Ser. 78 012057
[3] IceCube Monitoring Utility. Retrieved March 24, 2022. https://github.com/WIPACrepo/monitoring-scripts/blob/master/condor_utils.py#L436
[4] Larry Smarr et al. 2018. The Pacific Research Platform: Making High-Speed Networking a Reality for the Scientist. PEARC '18: Proceedings of the Practice and Experience on Advanced Research Computing. 29 pp 1–8. https://doi.org/10.1145/3219104.3219108
[5] D. Chirkin. 2013. Photon tracking with GPUs in IceCube. Nucl. Inst. Methods Phys. Res. Sec. A. 725, 141–143. https://doi.org/10.1016/j.nima.2012.11.170
[6] I. Sfiligoi, D. C. Bradley, B. Holzman, P. Mhashilkar, S. Padhi and F. Wurthwein. 2009. The Pilot Way to Grid Resources Using glideinWMS," 2009 WRI World Congress on Computer Science and Information Engineering pp. 428-432. https://doi.org/10.1109/CSIE.2009.950
[7] NVIDIA Multi-Instance GPU User Guide. Retrieved March 24, 2022. https://docs.nvidia.com/datacenter/tesla/mig-user-guide/
[8] Google Cloud GPU Pricing. Retrieved March 24, 2022. https://cloud.google.com/compute/gpus-pricing
[9] I. Sfiligoi, T. Martin, B. P. Bockelman, D. C. Bradley, and F. Würthwein. 2014. Minimizing draining waste through extending the lifetime of pilot jobs in Grid environments. J. Phys.: Conf. Ser. 513 032089